\documentclass[article,11pt]{article}

\pdfoutput=1

\usepackage{jheppub}
\usepackage{caption}
\usepackage{subcaption}
\usepackage{amssymb}
\usepackage{amsbsy}
\usepackage{amsfonts}
\usepackage{amsmath}
\usepackage{epsfig}
\usepackage{cancel}
\usepackage{slashed}
\usepackage{array}
\usepackage{mathtools}

\allowdisplaybreaks

\newcolumntype{L}[1]{>{\raggedright\let\newline\\\arraybackslash\hspace{0pt}}m{#1}}
\newcolumntype{C}[1]{>{\centering\let\newline\\\arraybackslash\hspace{0pt}}m{#1}}
\newcolumntype{R}[1]{>{\raggedleft\let\newline\\\arraybackslash\hspace{0pt}}m{#1}}
\newcommand\myeq{\stackrel{\mathclap{B_\mu \to B_\mu^\text{max}}}{\sim\;\;}}

\title{The spontaneous $\mathbb{Z}_2$ breaking Twin Higgs}

\author[a]{Hugues Beauchesne}
\author[a]{Kevin Earl}
\author[a]{Thomas~Gr\'egoire}

\emailAdd{HuguesBeauchesne@cmail.carleton.ca}
\emailAdd{KevinEarl@cmail.carleton.ca}
\emailAdd{gregoire@physics.carleton.ca}

\affiliation[a]{Ottawa-Carleton Institute for Physics, Department of Physics, Carleton University 1125 Colonel By Drive, Ottawa, K1S 5B6 Canada}
\abstract{The Twin Higgs model seeks to address the little hierarchy problem by making the Higgs a pseudo-Goldstone of a global $SU(4)$ symmetry that is spontaneously broken to $SU(3)$. Gauge and Yukawa couplings, which explicitly break $SU(4)$, enjoy a discrete $\mathbb{Z}_2$ symmetry that accidentally maintains $SU(4)$ at the quadratic level and therefore keeps the Higgs light. Contrary to most beyond the Standard Model theories, the quadratically divergent corrections to the Higgs mass are cancelled by a mirror sector, which is uncharged under the Standard Model groups. However, the Twin Higgs with an exact $\mathbb{Z}_2$ symmetry leads to equal vevs in the Standard Model and mirror sectors, which is phenomenologically unviable. An explicit $\mathbb{Z}_2$ breaking potential must then be introduced and tuned against the $SU(4)$ breaking terms to produce a hierarchy of vevs between the two sectors. This leads to a moderate but non-negligible tuning. We propose a model to alleviate this tuning, without the need for an explicit $\mathbb{Z}_2$ breaking sector. The model consists of two $SU(4)$ fundamental Higgses, one whose vacuum preserves $\mathbb{Z}_2$ and one whose vacuum breaks it. As the interactions between the two Higgses are turned on, the $\mathbb{Z}_2$ breaking is transmitted from the broken to the unbroken sector and a small hierarchy of vevs is naturally produced. The presence of an effective tadpole and feedback between the two Higgses lead to a sizable improvement of the tuning. The resulting Higgs boson is naturally very Standard Model like.
}

\begin{document}

\maketitle

\section{Introduction}
One of the goals of beyond the Standard Model (BSM) physics is to stabilize the hierarchy between the electroweak and Planck scale. To this end, most BSM models introduce partners that cancel the quadratic divergent corrections to the Higgs mass. These partners are generally assumed to be charged under the Standard Model (SM) groups. Unfortunately, the lack of discovery of new particles in Run-1 of the LHC has put strong constraints on these partners and further accentuates the little hierarchy problem \cite{Barbieri:2000gf}. One way to avoid this problem is {\it neutral naturalness}, the idea that partners are not charged under the SM groups. Perhaps the best example of this is Twin Higgs \cite{Chacko:2005pe} (see \cite{Barbieri:2005ri,Burdman:2006tz,Cai:2008au,Burdman:2008ek,Poland:2008ev,Batra:2008jy,Craig:2014aea,Craig:2014roa,Burdman:2014zta,Barbieri:2015lqa,Low:2015nqa,PhysRevD.92.055034,Craig:2015xla,PhysRevLett.115.121801,Farina:2015uea,Batell:2015aha,Craig:2015pha,Curtin:2015fna,Curtin:2015bka} for related work). This model rests on a global $SU(4)$ which is  broken spontaneously to $SU(3)$ at a scale $f$, leading to a set of Goldstone bosons. The $SU(4)$ is explicitly broken by gauging a $SU(2)_A \times SU(2)_B$ subgroup (with $SU(2)_A$ being identified with the SM $SU(2)$ and $SU(2)_B$ a similar symmetry of a mirror sector) and by adding Yukawa couplings. In principle, this breaking would give a mass of order $f$ to the Goldstone bosons. Remarkably, imposing a $\mathbb{Z}_2$ symmetry between the two sectors ensures that the theory is still $SU(4)$ invariant at the quadratic level, leading to a light pseudo-Goldstone Higgs. A soft $\mathbb{Z}_2$ breaking is however needed to obtain a hierarchy of vacuum expectation values (vev) between the Standard Model Higgs and the mirror sector Higgs \cite{Chacko:2005pe,Barbieri:2005ri}.\footnote{See section \ref{sSec:TH} for more details}

Despite its success, even the Twin Higgs is not free from tuning. A moderate amount of tuning between the $\mathbb{Z}_2$ and the $SU(4)$ breaking sectors is needed to push the cutoff beyond experimental constraints. Various attempts at addressing this issue can be found in the literature. Reference \cite{Chacko:2005vw} tries to do so in the context of a two Higgs doublet model with misaligned vevs. In \cite{Chang:2006ra}, the issue is addressed in a supersymmetric (SUSY) UV completion by introducing Dirac gauginos \cite{Fox:2002bu}. Finally, \cite{Falkowski:2006qq} also addresses the supersymmetric completion, but by forcing $\tan\beta=1$ in the mirror sector. Both of these models try to remove the D-term quartics which are a source of tuning in supersymmetric versions of the Twin Higgs.  One thing all of these models have in common is an explicit $\mathbb{Z}_2$ breaking.

In this article, we propose a novel approach to improving the tuning in Twin Higgs, which is based on spontaneous breaking of the $\mathbb{Z}_2$ symmetry. The proposed model includes two Higgses in the fundamental representation of a $SU(4)$ global symmetry. As in the original Twin Higgs model, a $SU(2)_A \times SU(2)_B$ subgroup is gauged and a $\mathbb{Z}_2$ symmetry is imposed between the two sectors. We take the vacuum of the first Higgs to preserve $\mathbb{Z}_2$, while the other breaks it spontaneously. A bilinear term containing the two Higgses is added (similar to  the $B_\mu$ term of the MSSM) and the $\mathbb{Z}_2$ breaking is transmitted from the broken to the unbroken sector. This naturally produces a hierarchy between the vevs of the SM sector Higgses and those of the mirror sector. The $B_\mu$ term acts as an effective tadpole and no explicit $\mathbb{Z}_2$ breaking is necessary. The presence of this effective tadpole and feedback between the two Higgses lead to less tuning than the original Twin Higgs. The resulting Higgs boson is naturally very SM-like.

The article is organized as follows. We begin by summarizing the original Twin Higgs model to isolate the origin of the tuning and obtain results that will make comparisons with our model easier. Our model is then presented in details. An analysis of the radiative corrections follows. A detailed analysis of the tuning of the model compared to the original Twin Higgs is then performed. Finally, a few concluding remarks including possible UV completions are presented.

\section{The model}\label{Sec:Model}

\subsection{The original Twin Higgs}\label{sSec:TH}
To put the problem our model attempts to solve in context and to establish our notation, we summarize the Twin Higgs model. We follow closely \cite{Chacko:2005pe}. Assume a complex scalar field $H$ which is a fundamental of a global $SU(4)$. Its potential can be written as
\begin{equation}\label{Eq:TH:VSU4}
	V_{SU(4)}(H)=-\mu^2H^{\dagger}H+\lambda(H^{\dagger}H)^2.
\end{equation}
The potential exhibits spontaneous symmetry breaking of $SU(4)\to SU(3)$. This leads to $\langle H\rangle\equiv f=\mu/\sqrt{2\lambda}$ and 7 Goldstone bosons. The SM-like Higgs doublet is associated to 4 of these Goldstone bosons and is at this stage massless.

The $SU(4)$ is then explicitly broken by gauging one of its $SU(2)_A \times SU(2)_B$ subgroups. The field $H$ is now divided into fundamentals of $SU(2)_A$ and $SU(2)_B$ as $H=(H_A,H_B)$. The $A$ sector is conventionally associated to the Standard Model and the $B$ sector to the mirror sector. The leading correction to the potential introduced by gauging the $SU(2)$'s is
\begin{equation}\label{Eq:TH:VSU4bquad}
	\Delta V(H)=\frac{9g_A^2 \Lambda^2}{64\pi^2}H_A^{\dagger}H_A+\frac{9g_B^2 \Lambda^2}{64\pi^2}H_B^{\dagger}H_B,
\end{equation}
where $g_A$ and $g_B$ are the gauge coupling constants of $SU(2)_A$ and $SU(2)_B$ respectively and $\Lambda$ is the cutoff of the theory. If a $\mathbb{Z}_2$ symmetry is imposed between the $A$ and $B$ sector, $g_A=g_B\equiv g$ and $\Delta V(H)$ accidentally respects the original $SU(4)$ symmetry. The Goldstone bosons therefore do not acquire any mass from \ref{Eq:TH:VSU4bquad}. Alternatively, one can then consider \ref{Eq:TH:VSU4bquad} as simply a correction to $\mu^2$. $SU(4)$ will however be broken by terms of the form $\kappa (|H_A|^4+|H_B|^4)$, where $\kappa$ is of order $g^2/16\pi^2\ln{(\Lambda/f)}$. These logarithmic divergences can be reabsorbed in $\lambda$ and a $SU(4)$ breaking potential of the form
\begin{equation}\label{Eq:TH:VSU4b}
	V_{\cancel{SU(4)}}(H)=\alpha H_A^{\dagger}H_A H_B^{\dagger}H_B.
\end{equation}
A similar story holds for the top Yukawa coupling. A $\mathbb{Z}_2$ symmetry is imposed on this sector by adding a {\lq mirror top\rq} which is not charged under the SM groups, but which couples to $H_B$ in exactly the same way in which the SM top couples to $H_A$. 

The total potential at this point is the sum of \ref{Eq:TH:VSU4} and \ref{Eq:TH:VSU4b}. The end result is that, of the original 7 Golstone bosons, 6 will remain massless and be eaten by massive gauge bosons and the one left over will be a light pseudo-Goldstone boson that can be associated to the 125 GeV Higgs. Since $\alpha$ is the only term in the potential that breaks $SU(4)$, it can naturally be smaller than $\lambda$, which is what we assume. This insures that the Higgs remains light even for relatively large $f$. 

The symmetry breaking structure is controlled by the sign of $\alpha$ \cite{Barbieri:2005ri}. If $\alpha<0$, the minimum preserves $\mathbb{Z}_2$ and $\langle H_A\rangle=\langle H_B\rangle=\mu/\sqrt{4\lambda+\alpha}\approx 174$ GeV. This is the sign of $\alpha$ assumed in the original Twin Higgs model. The fact that $\langle H_A\rangle=\langle H_B\rangle$ leads to the Standard Model Higgs strongly mixing with the mirror sector Higgs and results in large deviations of the Higgs measurements \cite{Barbieri:2005ri}. It also means that $f$ is only slightly above the electroweak scale. The energy scale $\sim 4 \pi f$, at which new physics needs to appear to avoid fine-tuning, is then not much larger than in the Standard Model.  These issues are easily resolved by aligning the vev closer to the $B$ sector, thereby allowing for a larger $f$ while preserving $\langle H_A\rangle=174$ GeV. This can be done via an explicit soft $\mathbb{Z}_2$ breaking potential of the form
\begin{equation}\label{Eq:TH:Z2b}
	V_{\cancel{\mathbb{Z}_2}}(H)=\Delta m^2 H_A^{\dagger}H_A.
\end{equation}
The parameter $\Delta m^2$ can naturally be small as it is the only term that explicitly breaks $\mathbb{Z}_2$. The potential can be minimized by using the following parametrization of the relevant parts of $H$
\begin{equation}\label{Eq:TH:Decomposition}
	H=f\begin{pmatrix} 0 \\ \sin\theta \\ 0 \\ \cos\theta\end{pmatrix},
\end{equation}
with $\theta$ being $\pi/4$ when $\Delta m^2$ is 0. The potential is minimized for a value of $f$ of
\begin{equation}\label{eq:US:RRRR}
	f^2=\frac{2\mu^2-\Delta m^2}{4\lambda+\alpha},
\end{equation}
while minimizing the potential with respect to $\theta$ gives the following equation
\begin{equation}\label{Eq:TH:Minimization}
	\alpha f^4 \sin 4\theta+4\Delta m^2 f^2 \sin\theta\cos\theta=0.
\end{equation}
This equation only yields non-zero $\theta$ for $\Delta m^2$ below a maximal value. Thus, we define $\Delta m^2_\text{max}$ as the largest value of $\Delta m^2$ for which there is still electroweak symmetry breaking in the $A$ sector. It can be found by rewriting \ref{Eq:TH:Minimization} as
\begin{equation}\label{Eq:TH:f1}
	F_1(\theta)\equiv\frac{1}{4}\frac{\sin 4\theta}{\sin\theta\cos\theta}= \frac{\Delta m^2}{(-\alpha f^2)}\approx\frac{\Delta m^2}{\Delta m^2_{\text{max}}},
\end{equation}
where the last relation holds in the limit of small $\alpha$ and $\Delta m^2_\text{max}$ is given by the exact relation
\begin{equation}\label{Eq:TH:Deltam2max}
	\Delta m^2_\text{max}=-\frac{\alpha \mu^2}{2\lambda}.
\end{equation}
The solution to  equation \ref{Eq:TH:Minimization} is
\begin{equation}\label{Eq:TH:v2f2}
	\sin^2\theta=\frac{v^2}{f^2}=\frac{1}{2}\left(1-\frac{\Delta m^2}{(-\alpha f^2)}\right)\approx\frac{1}{2}\left(1-\frac{\Delta m^2}{\Delta m^2_{\text{max}}}\right),
\end{equation}
where $v$ is the SM Higgs vev of 174 GeV. Requesting a large $f$ implies a tuning between the $SU(4)$ breaking and the $\mathbb{Z}_2$ breaking potentials. This is reflected in \ref{Eq:TH:v2f2} by the last term on the right needing to be close to 1.

Alternatively, one can take $\alpha>0$. The $\mathbb{Z}_2$ symmetry is then spontaneously broken and the system falls in one of the two minima at $\langle H_A\rangle=\mu/\sqrt{2\lambda}$ and $\langle H_B\rangle=0$ or $\langle H_A\rangle=0$ and $\langle H_B\rangle=\mu/\sqrt{2\lambda}$. However, the vev must be taken to fall in the SM sector and this leads to a massless mirror sector. This proves to be unviable for cosmological reasons \cite{Barbieri:2005ri}. The potential must then be modified in a way similar to \ref{Eq:TH:Z2b} to prevent the minimum from being in one sector only. Unfortunately, a quick inspection shows that no term that only breaks $\mathbb{Z}_2$ softly and respects gauge invariance can do so. The term of equation \ref{Eq:TH:Z2b} does not solve this problem, as equation \ref{Eq:TH:Minimization} is satisfied by a $\theta$ of 0 for all values of $\Delta m^2$. The case of $\alpha>0$ therefore poses serious issues.

\subsection{Spontaneous $\mathbb{Z}_2$ breaking}\label{sSec:USmodel}
In the last section, part of the problem in the $\alpha>0$ case was that $H$ was the only scalar with gauge charges. This forced all terms in the potential to be an even power of $H$ and forbade tadpole terms, which could have potentially prevented the vev from falling in one sector only. The inclusion of a second Higgs field can solve this problem by including a term linear in both fields which acts as an effective tadpole for $H$ (see \cite{Galloway:2013dma} for a similar idea in a context unrelated to $\mathbb{Z}_2$ breaking or the Twin Higgs). In addition, the $\mathbb{Z}_2$ breaking soft term for $\alpha<0$ also needed to be quadratic in $H$. It is possible that a similar term with a lower power of $H$ could potentially produce the same hierarchy of vevs while requiring less tuning. Again, a term linear in $H$ and another Higgs can do this. Taking these considerations into account, our model includes two fundamentals of $SU(4)$, $H_1=(H_{1A},H_{1B})$ and $H_2=(H_{2A},H_{2B})$, that are gauged as in Twin Higgs and which interact with each other to create a hierarchy of vevs. We take the minimum of $H_1$ to preserve $\mathbb{Z}_2$ and that of $H_2$ to break it. It is the interaction between $H_1$ and  $H_2$ that transmits the $\mathbb{Z}_2$ breaking to $H_1$ and there is no need for an explicit $\mathbb{Z}_2$ breaking. We explain the finer details below.

\subsubsection{Potential and vevs}\label{ssSec:PotentialandVevs}
As a starting point, we write down the potential for $H_1$ by itself
\begin{equation}\label{Eq:US:H1}
	V_{H_1}(H_1)=-\mu_1^2H_1^{\dagger}H_1+\lambda_1(H_1^{\dagger}H_1)^2+\alpha_1 {H_{1A}}^{\dagger}H_{1A} {H_{1B}}^{\dagger}H_{1B}
\end{equation}
and assume $\alpha_1<0$, which means that the vacuum preserves $\mathbb{Z}_2$. At this point, the pseudo-Goldstone boson from $H_1$ corresponds to the angular mode and is an equal admixture of the components of $H_{1A}$ and $H_{1B}$. Similarly, we write a potential for $H_2$ by itself
\begin{equation}\label{Eq:US:H2}
	V_{H_2}(H_2)=-\mu_2^2H_2^{\dagger}H_2+\lambda_2(H_2^{\dagger}H_2)^2+\alpha_2 {H_{2A}}^{\dagger}H_{2A} {H_{2B}}^{\dagger}H_{2B}
\end{equation}
and this time with $\alpha_2>0$, meaning that the vacuum breaks $\mathbb{Z}_2$ in this case. We take the vev to fall in the $B$ sector by convention, as the vev falling in the other sector would just mean a relabelling of $B$ as the SM and $A$ as the mirror sector. The pseudo-Goldstone boson again corresponds to the angular mode. This time however, the position of the minimum means that the pseudo-Goldstone boson is purely a component of $H_{2A}$.

\begin{figure}[t!] 
	\centering 
	\includegraphics[width=0.6\textwidth, bb = 0 0 411 308]{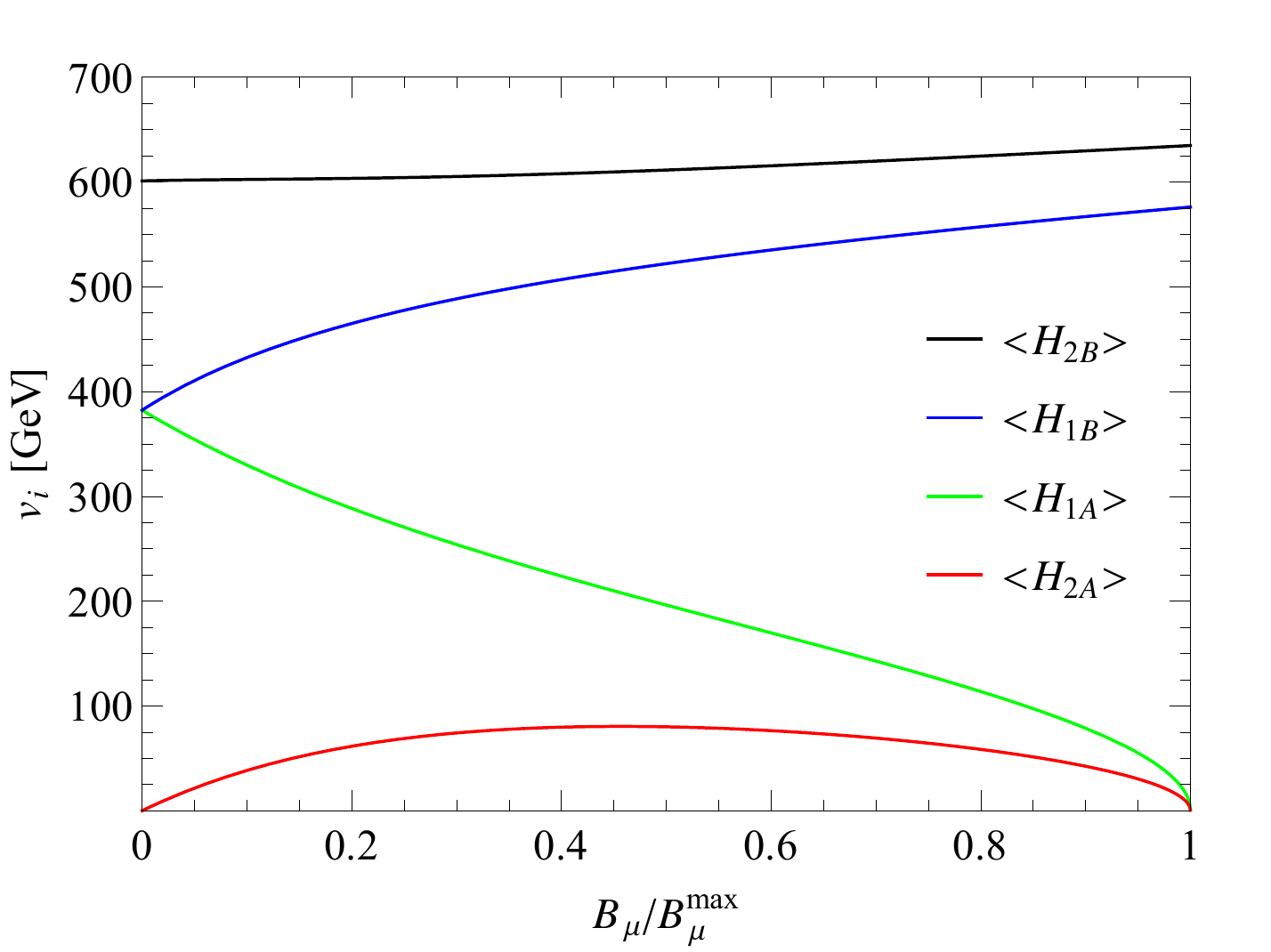} 
	\caption{Example of the different vevs as a function of $B_\mu / B_\mu^\text{max}$. The parameters are $\mu_1=750$ GeV, $\mu_2=850$ GeV, $\alpha_1=-0.15$, $\alpha_2=0.2$ and $\lambda_1=\lambda_2=1$.}
	\label{Fig:vevs}
\end{figure}

The interaction between these two fields is then codified by the following Lagrangian
\begin{equation}\label{Eq:US:Bmu}
	V_{H_1H_2}(H_1,H_2)=-B_\mu H_1^\dagger H_2+ \text{h.c.}.
\end{equation}
We note that it is technically natural to have $B_\mu$ small as it breaks a Peccei-Quinn symmetry. For $B_\mu$ small and greater than zero, \ref{Eq:US:Bmu} serves essentially two purposes. First, the part $H_{1B}^{\dagger}H_{2B}$ serves as an effective tadpole for $H_{1B}$. It therefore pushes the vev of $H_1$ toward the $B$ sector, as desired. Second, the part $H_{1A}^{\dagger}H_{2A}$ serves as an effective tadpole for $H_{2A}$. It accordingly provides a small positive $A$ component to the vev of $H_2$. As $B_\mu$ increases, non-linear effects and feedback between the different terms become important. An example of the different vevs is shown in figure \ref{Fig:vevs}.

\subsubsection{Small $\alpha_i$'s approximation}\label{ssSec:Smallalphaaprrox}
To gain a better understanding of the interactions between $H_1$ and $H_2$, we decompose them in a similar way to \ref{Eq:TH:Decomposition} and take the limit of small $\alpha_i$'s. As will be made clear in equation \ref{Eq:US:BmuMax}, $B_\mu$ will be a factor of $\alpha_1/\lambda_1$ smaller than the $\mu_i^2$'s in the physically viable and natural region of parameter space. We therefore assume it to be small. In general, all approximations will be valid up to $\mathcal{O}(\alpha_i/\lambda_i)$. The decomposition of the Higgses is
\begin{equation}\label{Eq:US:Decomposition}
	H_1=f_1\begin{pmatrix} 0 \\ \sin\theta_1 \\ 0 \\ \cos\theta_1\end{pmatrix} \;\;\;\;\;\;\;\;\;\;\; H_2=f_2\begin{pmatrix} 0 \\ \sin\theta_2 \\ 0 \\ \cos\theta_2\end{pmatrix},
\end{equation}
where $f_1\approx\mu_1/\sqrt{2\lambda_1}$ and $f_2\approx\mu_2/\sqrt{2\lambda_2}$. The minimization of the potential with respect to the angles leads to the set of equations
\begin{equation}\label{Eq:US:Minimization}
	\begin{aligned}
		&\alpha_1 f_1^4\sin 4\theta_1 + 4B_\mu f_1 f_2 \sin(\theta_1-\theta_2)=0 \\
		&\alpha_2 f_2^4\sin 4\theta_2 - 4B_\mu f_1 f_2 \sin(\theta_1-\theta_2)=0.
	\end{aligned}
\end{equation}
When $B_\mu=0$, the minimum is located at $\theta_1=\pi/4$ and $\theta_2=0$. In the general case, adding both equations leads to
\begin{equation}\label{Eq:US:AngleRelation}
	\sin 4\theta_2=\Omega\sin 4\theta_1,
\end{equation}
where $\Omega$ is a constant in the small $\alpha$ approximation and is defined by
\begin{equation}\label{Eq:US:Omega}
	\Omega\equiv-\frac{\alpha_1}{\alpha_2}\left(\frac{f_1}{f_2}\right)^4.
\end{equation}
First, consider $\Omega < 1$. Increasing $B_\mu$ will make $\theta_1$ pass from $\pi/4$ to 0. The angle $\theta_2$ starts by increasing but decreases once $\theta_1$ drops below $\pi/8$. Eventually, both angles settle at 0. When $\Omega > 1$, this behavior is reversed. Increasing $B_\mu$ will make $\theta_2$ pass from 0 to $\pi/4$. The angle $\theta_1$ decreases until $\theta_2$ reaches $\pi/8$, but increases afterward. Both angles ultimately settle to $\pi/4$. This behavior is not bad in itself as it can still lead to a small hierarchy, but obtaining a large one proves to be impossible. Taking these considerations into account, we focus on the domain where $\Omega < 1$. 

Analogous to the Twin Higgs case, we define $B_\mu^\text{max}$  as the largest value of $B_\mu$ for which there is still electroweak symmetry breaking in the $A$ sector. The first equation of \ref{Eq:US:Minimization} can then be rewritten as
\begin{equation}\label{Eq:US:f2}
	F_2(\theta_1,\Omega)\equiv\frac{(1-\Omega)}{4}\frac{\sin 4\theta_1}{\sin(\theta_1-\theta_2)}= \frac{B_\mu}{\left(-\frac{\alpha_1 f_1^3}{f_2(1-\Omega)}\right)} \approx \frac{B_\mu}{B_\mu^\text{max}},
\end{equation}
where $\theta_2$ is related to $\theta_1$ by equation \ref{Eq:US:AngleRelation}. In the small $\alpha_i$'s approximation, $B_\mu^\text{max}$ is then
\begin{equation}\label{Eq:US:BmuMax}
	B_\mu^\text{max}\approx-\frac{\alpha_1 f_1^3}{f_2(1-\Omega)}.
\end{equation}
While it is hard to solve \ref{Eq:US:f2} for $\theta_1$, it is easy to see that small values of $\theta_1$ require $B_{\mu}$ to be close to $B_\mu^\text{max}$. This is similar to the Twin Higgs case where $\Delta m^2$ needed to be close to $\Delta m^2_\text{max}$ to obtain a small ratio of vevs. 

\begin{figure}[t!] 
	\centering 
	\includegraphics[width=0.6\textwidth, bb = 0 0 411 316]{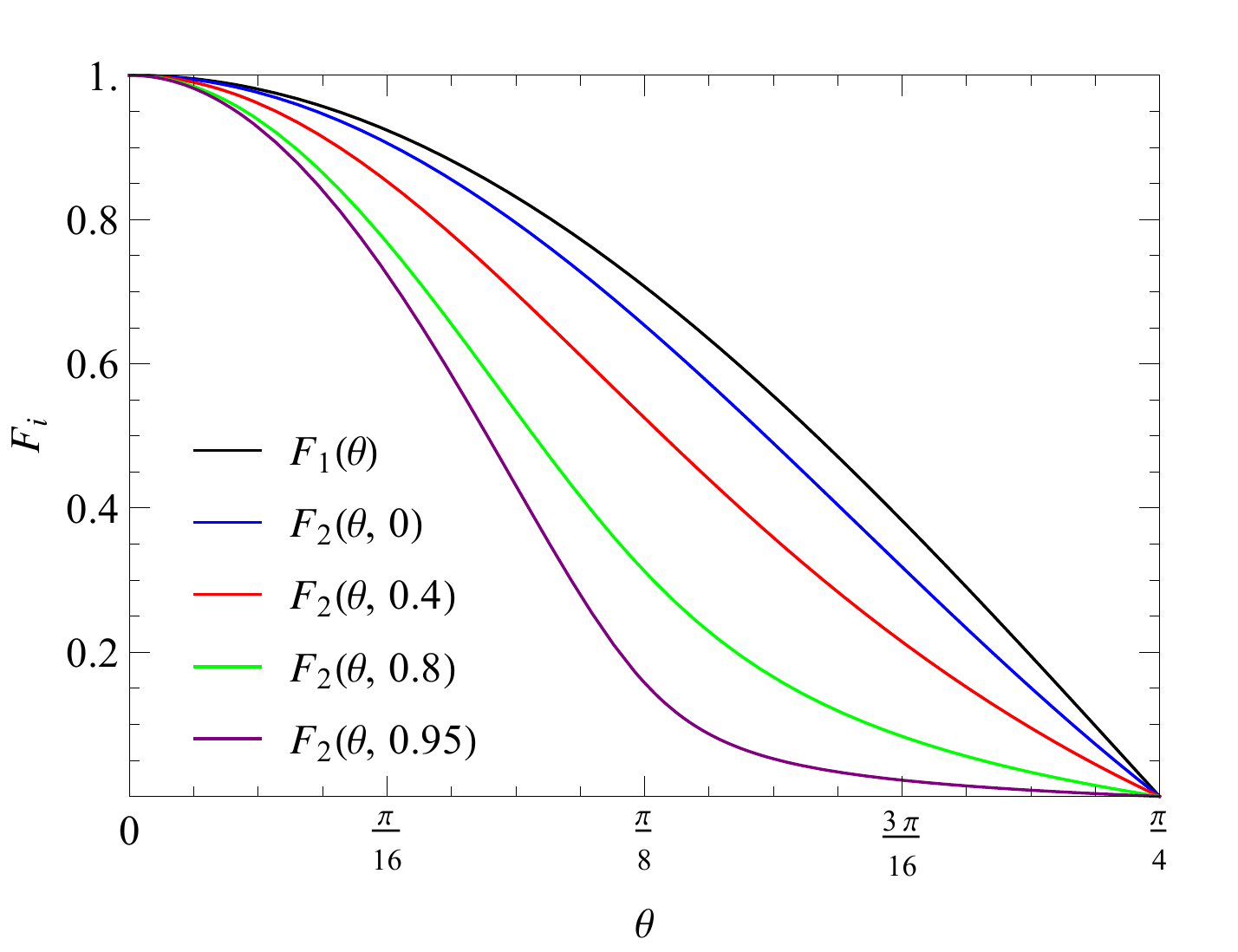} 
	\caption{$F_1(\theta)$ and $F_2(\theta,\Omega)$ for different values of $\Omega$.}
	\label{Fig:F1F2}
\end{figure}

We can compare the two theories by looking at $F_1(\theta)$ versus $F_2(\theta,\Omega)$ which are plotted in figure \ref{Fig:F1F2} for different values of $\Omega$ between 0 and 1.
When $0 < \theta < \pi/4$, $F_2(\theta,\Omega)$ is always smaller than $F_1(\theta)$. This means that, for the same angle, our model doesn't require $B_\mu$ as close to $B_\mu^\text{max}$ as the Twin Higgs requires $\Delta m^2$ close to $\Delta m_\text{max}^2$. This translates to less tuning. In contrast to the Twin Higgs, one must keep in mind that for our model $\langle H_{1A} \rangle < v = 174$ GeV, as it is a two Higgs doublet model. As avoiding large tuning requires new physics near $\sim 4\pi f_1$, this suggests that for equivalent tuning and cutoff one must choose $\theta_1$ smaller than the equivalent angle in Twin Higgs. Fortunately, our model naturally leads to $\langle H_{1A}\rangle$ considerably larger than $\langle H_{2A}\rangle$. Thus, the difference is small and the argument about tuning remains valid.

\begin{figure}[t!] 
	\centering 
	\includegraphics[width=0.6\textwidth, bb = 0 0 411 307]{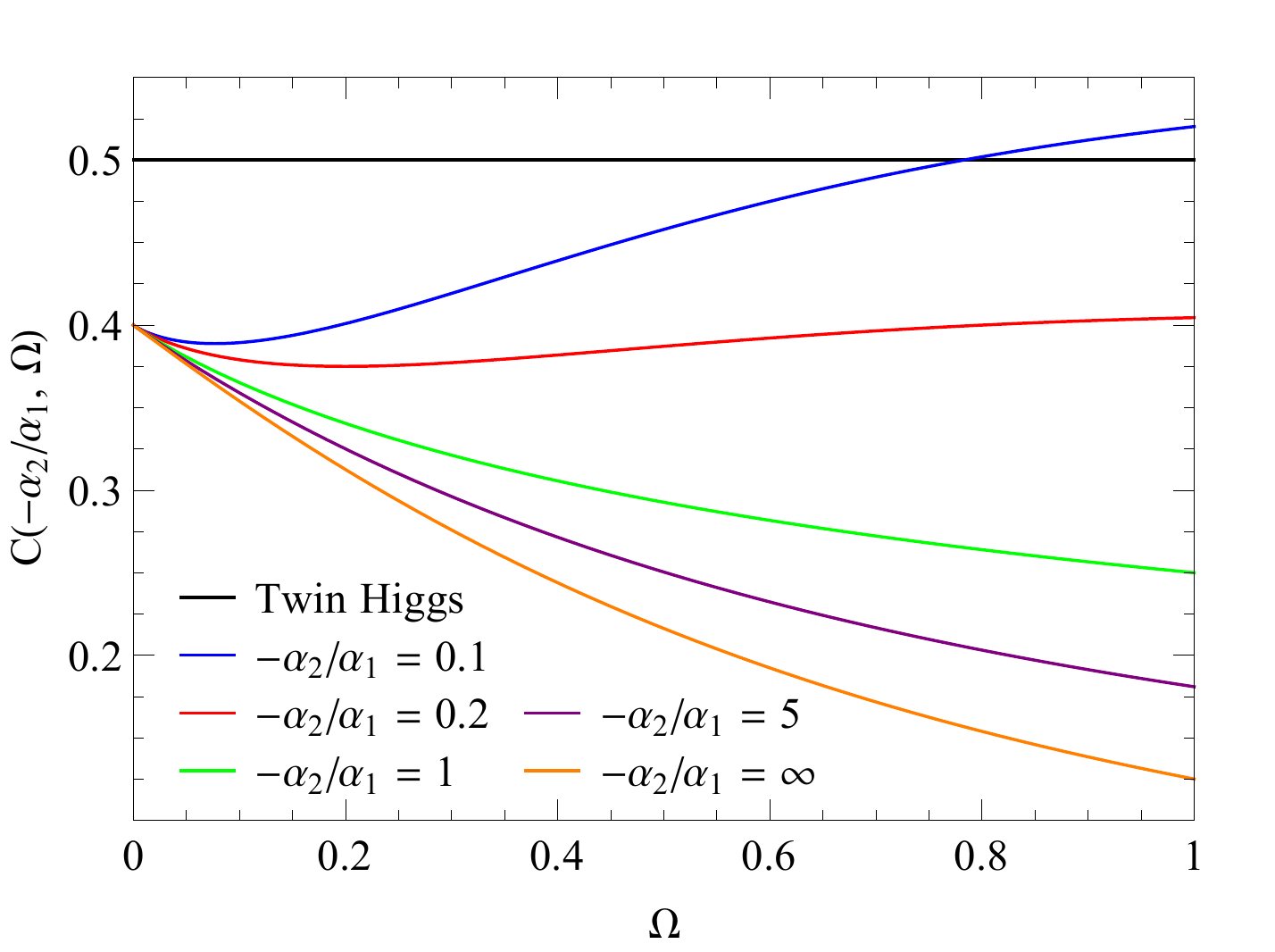} 
	\caption{$C\left(-\alpha_2/\alpha_1,\Omega\right)$ for various values for $-\alpha_2/\alpha_1$. Also shown is the corresponding value for Twin Higgs.}
	\label{Fig:small_angle}
\end{figure}

Further insight can be obtained by taking the small $\theta_1$ limit of \ref{Eq:US:f2}
\begin{equation}\label{Eq:US:theta12}
	\theta^2_1\approx \frac{3}{8}\frac{(B_\mu^\text{max}-B_\mu)}{(B_\mu^\text{max}+g(\Omega)B_\mu)} \;\;\;\; \myeq \;\;\; \frac{3}{8(1+g(\Omega))}\left(1-\frac{B_\mu}{B_\mu^\text{max}}\right),
\end{equation}
where
\begin{equation}\label{Eq:US:gOmega}
	g(\Omega)\equiv \frac{1}{16}(15\Omega^2+18\Omega-1).
\end{equation}
As mentioned above, a more appropriate quantity to make the comparison with the Twin Higgs is
\begin{equation}\label{Eq:US:v2f2smalltheta}
	\begin{aligned}
	\frac{v^2}{f_1^2} & \sim\frac{3}{8(1+g(\Omega))}\left(1+\left(-\frac{\alpha_2}{\alpha_1}\right)^{-1/2}\Omega^{3/2}\right)\left(1-\frac{B_\mu}{B_\mu^\text{max}}\right)\\
	& \equiv C\left(-\alpha_2/\alpha_1,\Omega\right)\left(1-\frac{B_\mu}{B_\mu^\text{max}}\right).
	\end{aligned}
\end{equation}
This is to be compared to \ref{Eq:TH:v2f2} which has a similar structure but with $C\left(-\alpha_2/\alpha_1,\Omega\right)$ replaced by $1/2$. Figure \ref{Fig:small_angle} shows $C\left(-\alpha_2/\alpha_1,\Omega\right)$ as a function of $\Omega$ for fixed values of $-\alpha_2/\alpha_1$. We see that, unless $-\alpha_2/\alpha_1$ is very small, $C\left(-\alpha_2/\alpha_1,\Omega\right)$ is smaller than $1/2$ for $\Omega$ in the whole range of 0 to 1. This shows that our model can easily obtain the same cutoff as the Twin Higgs for less tuning.

The improvement in the tuning can ultimately  be attributed to two sources. First, we can look at the limit of small $\Omega$, which means that $\theta_2$ is also small. This limit means that $H_2$ only serves as an effective tadpole and does not mix with $H_1$. The fact that the symmetry breaking is induced by an effective tadpole translates to \ref{Eq:US:f2} missing the factor of $\cos\theta$ present in \ref{Eq:TH:f1}. This by itself is enough to insure that $F_2(\theta,\Omega)$ be smaller than $F_1(\theta)$. Second, there is considerable feedback between $H_1$ and $H_2$ when $\Omega$ is close to 1. This translates to $\theta_2$ and $1-\Omega$ appearing in \ref{Eq:US:f2}. The presence of these terms further decreases $F_2(\theta,\Omega)$, as is clearly shown in figure \ref{Fig:F1F2}. Obviously, taking $\Omega$ close to 1 is a tuning in itself, though certainly not large enough to spoil our results, and we take this into account in section \ref{Sec:FineTuning}.

\subsubsection{Additional properties}\label{ssSec:AdditionalProperties}
A few additional properties of the model are worth mentioning. The first one is that the behavior of figure \ref{Fig:vevs} can differ outside of the region of parameter space considered up to now. The case of $\Omega > 1$ mentioned above is an example. Even when $\Omega < 1$, the vevs can act differently if the $\alpha_i$'s or $B_\mu$ are large. In particular, it is possible to choose parameters such that the vevs of the $A$ sector start like those of figure \ref{Fig:vevs} but fail to reach 0. It is also possible for the vevs of the $A$ sector to be 0 for an interval of $B_\mu$ but then become non-zero again for very large $B_\mu$. We therefore define more precisely $B_\mu^\text{max}$ as the smallest positive value of $B_\mu$ for which the vevs of the $A$ sector are zero. Fortunately, a sufficient condition for $B_\mu^\text{max}$ to exist, which is that the vevs of the $A$ sector settle to 0 for large $B_\mu$, is easily satisfied and given by
\begin{equation}\label{Eq:US:BmuMaxExCondition}
	\frac{\alpha_1}{\lambda_1}+\frac{\alpha_2}{\lambda_2}+\frac{\alpha_1\alpha_2}{2\lambda_1\lambda_2}>0.
\end{equation}
When this relation is close to being satisfied but not quite, it is possible that the system falls in the scenario where the vevs in the $A$ sector are 0 for an interval but become positive again for large $B_\mu$. This relation comes from looking at the limit of large $B_\mu$, where the $\mu_i^2$'s can be ignored. In this case, setting $\alpha_2$ to 0 will result in the potential being minimized for both $\theta_i$'s being $\pi/4$. Increasing $\alpha_2$ while keeping the other parameters fixed will cause both angles to eventually move toward 0. The angles will settle to 0 (which is always an extremum) when this point becomes a minimum, which happen when  \ref{Eq:US:BmuMaxExCondition} is satisfied. The vevs of the $A$ sector will then be 0 for large enough $B_\mu$ and it is therefore sufficient for $B_\mu^\text{max}$ to exist. 

\begin{figure}[t!] 
	\centering 
	\includegraphics[width=0.6\textwidth, bb = 0 0 411 313]{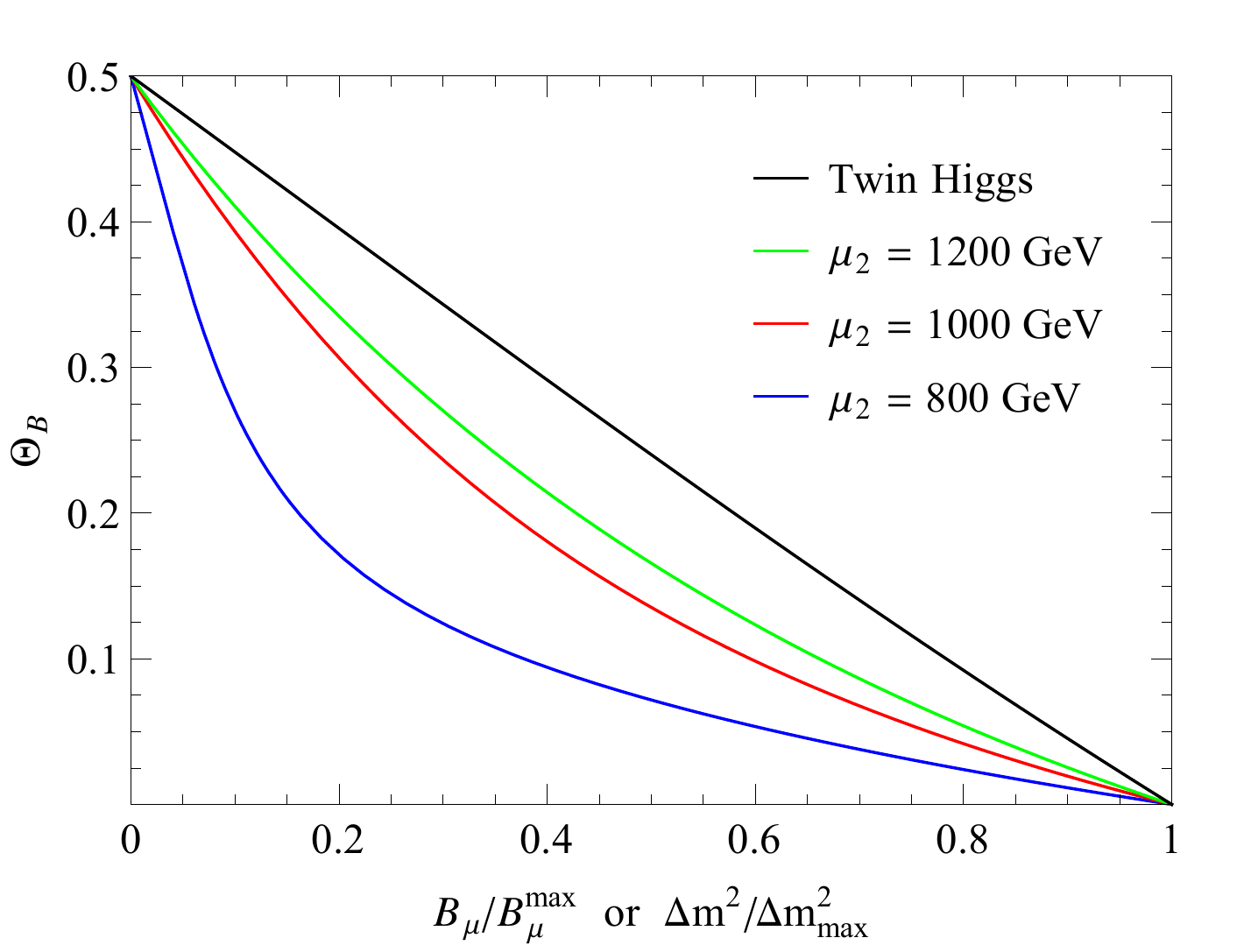} 
	\caption{Example of $\Theta_B$ for the Twin Higgs and spontaneous $\mathbb{Z}_2$ breaking model with different values of $\mu_2$. The parameters for the Twin Higgs model are $\mu=750$ GeV, $\alpha=-0.15$ and $\lambda=1$. The parameters for the spontaneous $\mathbb{Z}_2$ breaking model are $\mu_1=750$ GeV, $\alpha_1=-0.15$, $\alpha_2=0.2$ and $\lambda_1=\lambda_2=1$.}
	\label{Fig:Blike}
\end{figure}

Also of importance is that when $B_\mu=0$ the pseudo-Goldstone boson from $H_1$ is an equal combination of the $A$ and $B$ sector, while the one from $H_2$ is purely in the $A$ sector. One would then expect that turning on $B_\mu$ would cause the resulting light pseudo-Golstone boson to be more $A$-like than in the equivalent case for Twin Higgs. This turns out to be the case. To see this, we decompose the lightest pseudo-Golstone as 
\begin{equation}\label{Eq:US:Mixing}
	h = a h_{1A} + b h_{2A} + c h_{1B}+ d h_{2B},
\end{equation}
where $h_{1A}$ is defined via $H_{1A}^0=(v_{1A}+(h_{1A}+iA_{1A})/\sqrt{2})$ and identically for the other $h_i$'s. The parameter $\Theta_B\equiv c^2+d^2$ represents a measure of how much the Higgs is $B$-like. A similar quantity can easily be defined for the Twin Higgs. The comparison for both models can be seen in figure \ref{Fig:Blike}. Note that the pseudo-Goldstone is most $A$-like for large mixing between $H_1$ and $H_2$. The price to pay for this is that constraints akin to those in the usual two Higgs doublets model become important. Fortunately, these constraints can easily be avoided, as the model naturally leads to a hierarchy between the vevs in the $A$ sector and fairly little mixing with mirror sector Higgses. Generally speaking, this means that our model will be better at avoiding constraints on Higgs couplings, though a full study of this is beyond the scope of this article.

The particle spectrum in the $A$ sector is the usual two Higgs doublet model one. Generically speaking, creating a small hierarchy will push the masses of the heavier Higgses up for a fixed value of the lightest Higgs. The constraints from heavy scalar searches can therefore be easily avoided.

Another point worth mentioning is that the potential we wrote down does not contain all possible $\mathbb{Z}_2$ preserving terms. We verified that these extra terms do not affect the qualitative behavior of the system, as long as they are not much bigger than the terms already included. Even small explicit $\mathbb{Z}_2$ breaking terms do not affect the qualitative behavior of the system. Unless stated otherwise, such terms will be ignored from now on to avoid obscuring the analysis. 

\subsection{Radiative corrections}\label{sSec:Radiative}

In this section, we present the one-loop leading radiative corrections for both the Twin Higgs and our model. Due to the similarities between both models, the radiative corrections are nearly identical for the two. The main differences result from the Twin Higgs only having a single $SU(4)$ fundamental while our model contains two. These results are also similar to the radiative corrections given in \cite{Chacko:2005vw}, another Twin Higgs model with two $SU(4)$ fundamentals. The differences between their radiative corrections and ours follow from different forms of the quartic interactions. 

To compute the radiative corrections, it is necessary to specify how the top couples to the different Higgses. In the Twin Higgs, this is encoded in the Lagrangian
\begin{equation}\label{eq:TH:TopY}
\mathcal{L}_\text{top} = -y_t (\overline{q}_A \tilde{H}_{A} t^c_A + \overline{q}_B \tilde{H}_{B} t^c_B)+\text{h.c.},
\end{equation}
where the $B$ sector quarks $q_B$ and $t^c_B$ do not carry Standard Model color and the tilde notation stands for $\tilde{H}=i\sigma_2H^*$. The other Yukawa couplings can be safely ignored. The leading radiative corrections to the parameters of the Twin Higgs are then
\begin{align}
	\delta \mu^2 &= \frac{1}{16\pi^2} \left(6 y_t^2 - \frac{9}{4}g^2 - \frac{3}{4}{g'}^2 - 10\lambda - 2\alpha \right) \Lambda^2, \label{Eq:TH:rcmu2}\\
	\delta \lambda &= \frac{1}{16\pi^2} \left(6 y_t^4 -\frac{9}{8}g^4 - \frac{3}{4}g^2{g'}^2 - \frac{3}{8}{g'}^4 - 32\lambda^2 - 8\lambda \alpha - 2\alpha^2 \right) \ln \frac{\Lambda}{f}, \label{Eq:TH:rclambda}\\
	\delta \alpha &= \frac{1}{16\pi^2} \left(-12 y_t^4 + \frac{9}{4}g^4 + \frac{3}{2}g^2{g'}^2 + \frac{3}{4}{g'}^4 - 24 \lambda \alpha \right) \ln \frac{\Lambda}{f}, \label{Eq:TH:rcalpha}\\
	\delta \Delta m^2 &= \frac{1}{16\pi^2} \left(-4\lambda +4\alpha \right) \Delta m^2 \ln \frac{\Lambda}{f}, \label{Eq:US:rcDeltam2}
\end{align}
where $y_t$ is the top Yukawa coupling, $g$ and $g'$ are the SM gauge couplings and $\Lambda$ denotes the cutoff scale of the theory.

For our model, we must also specify how the top sector couples to the various Higgses. We choose the top to couple to $H_1$ only and to follow the structure of equation \ref{eq:TH:TopY}. The radiative corrections also depend on how the down-type quarks and the charged lepton couple to the Higgses, but the size of their Yukawa couplings makes these contributions irrelevant.

Another difference between our model and the Twin Higgs is that, in our case, radiative corrections also generate an additional operator of the form 
\begin{equation}\label{Eq:US:kappaTerm}
-\kappa (H_{1A}^{\dagger} H_{1A} H_{2A}^{\dagger} H_{2A} + H_{1B}^{\dagger} H_{1B} H_{2B}^{\dagger} H_{2B}).
\end{equation}
As mentioned above, the presence of a such a term does not modify qualitatively the behavior of the potential, as long as its coefficient is sufficiently small. We verified that this is the case for the operator of equation \ref{Eq:US:kappaTerm} with a coefficient of the size of its radiative correction. Even a considerably larger coefficient does not affect the behavior much. Because of this, we limit ourselves to writing down its radiative correction and ignore it afterward. The leading radiative corrections then take the form
\begin{align}
	\delta \mu_1^2 &= \frac{1}{16\pi^2} \left(6 y_t^2 - \frac{9}{4}g^2 - \frac{3}{4}{g'}^2 - 10\lambda_1 - 2\alpha_1 \right) \Lambda^2, \label{Eq:US:rcmu12}\\
	\delta \lambda_1 &= \frac{1}{16\pi^2} \left(6 y_t^4 -\frac{9}{8}g^4 - \frac{3}{4}g^2{g'}^2 - \frac{3}{8}{g'}^4 - 32\lambda_1^2 - 8\lambda_1 \alpha_1 - 2\alpha_1^2 \right) \ln \frac{\Lambda}{f_1}, \label{Eq:US:rclambda1}\\
	\delta \alpha_1 &= \frac{1}{16\pi^2} \left(-12 y_t^4 + \frac{9}{4}g^4 + \frac{3}{2}g^2{g'}^2 + \frac{3}{4}{g'}^4 - 24 \lambda_1 \alpha_1 \right) \ln \frac{\Lambda}{f_1}, \label{Eq:US:rcalpha1}\\
	\delta \mu_2^2 &= \frac{1}{16\pi^2} \left(-\frac{9}{4}g^2 - \frac{3}{4}{g'}^2 - 10\lambda_2 - 2\alpha_2 \right) \Lambda^2, \label{Eq:US:rcmu22}\\
	\delta \lambda_2 &= \frac{1}{16\pi^2} \left(-\frac{9}{8}g^4 - \frac{3}{4}g^2{g'}^2 - \frac{3}{8}{g'}^4 - 32\lambda_2^2 - 8\lambda_2 \alpha_2 - 2\alpha_2^2 \right) \ln \frac{\Lambda}{f_2}, \label{Eq:US:rclambda2}\\
	\delta \alpha_2 &= \frac{1}{16\pi^2} \left(\frac{9}{4}g^4 + \frac{3}{2}g^2{g'}^2 + \frac{3}{4}{g'}^4 - 24 \lambda_2 \alpha_2 \right) \ln \frac{\Lambda}{f_2}, \label{Eq:US:rcalpha2}\\
	\delta B_\mu &= 0, \label{Eq:US:rcbmu}\\
	\delta \kappa &= \frac{1}{16\pi^2} \left(-\frac{9}{4}g^4 - \frac{3}{2}g^2{g'}^2 - \frac{3}{4}{g'}^4 \right) \ln \frac{\Lambda}{f_1}. \label{Eq:US:rckappa}
\end{align}
For all radiative corrections presented above, we have neglected finite contributions.
\section{Numerical analysis of the fine-tuning}\label{Sec:FineTuning}
In this section, we seek to compare more precisely the fine-tuning of our model to the original Twin Higgs. For both models, the fine-tuning comes from requesting a small $v/f$. In the case of the Twin Higgs, one has to tune the $\mathbb{Z}_2$ breaking sector against the $SU(4)$ breaking sector. The tuning is evaluated in a similar way to \cite{Craig:2013fga} by defining
\begin{equation}\label{Eq:TH:Tuning}
	\Delta_{\text{TH}}=\left|\frac{\partial\ln(v^2/f^2)}{\partial\ln \Delta m^2}\right|.
\end{equation}
The tuning is then $\Delta_{\text{TH}}^{-1}$. There are however a number of constraints that need to be satisfied. The vev $v$ and the mass of the lightest Higgs must be adjusted to their correct values, which we take to be 174.10 GeV \cite{Agashe:2014kda} and 125.09 GeV \cite{Aad:2015zhl} respectively. In addition, $f/v$ must be large enough to avoid experimental constraints. Setting this ratio to a given value imposes an additional constraint. Alternatively, one can set the fine-tuning to a given number and be interested in $f/v$, which can be used to estimate the cutoff.

There are four parameters in the Twin Higgs potential: $\mu^2$, $\lambda$, $\alpha$ and $\Delta m^2$. Matching $v$ and the mass of the Higgs with their respective values sets two parameters. Fixing $f/v$ or the tuning determines another one. We are therefore left with a single free parameter. For convenience sake, we take that parameter to be $\lambda$. We give two benchmarks. First, setting $\lambda=1$ and $f/v=3$ leads to a tuning of 27.7$\%$. Second, setting $\lambda=1$ and requesting a tuning of 20$\%$ leads to a $f/v$ of 3.42.

A similar measure of fine-tuning can be defined in our model, but a few differences need to be taken into account. First, $B_\mu$ plays a similar role to $\Delta m^2$. As explained in section \ref{sSec:USmodel}, one can obtain a very large ratio of vevs for a relatively small $B_\mu/B_\mu^{\text{max}}$, given a very large mixing of $H_1$ and $H_2$. This however requires a fine-tuning of the parameters of the second Higgs ($\mu_2^2$, $\lambda_2$ and $\alpha_2$) against those of the first. This tuning corresponds to $\Omega$ being close to 1 and needs to be taken into account. Second, there are simply more parameters in our case than in the original Twin Higgs. A measure that addresses all of these issues in a relatively fair manner is
\begin{equation}\label{Eq:US:Tuning}
	\Delta_{\text{Spontaneous}}=\text{Max}\left\{\left|\frac{\partial\ln(v^2/f_1^2)}{\partial\ln B_\mu}\right|,\left|\frac{\partial\ln(v^2/f_1^2)}{\partial\ln \mu_2^2}\right|,\left|\frac{\partial\ln(v^2/f_1^2)}{\partial\ln \lambda_2}\right|,\left|\frac{\partial\ln(v^2/f_1^2)}{\partial\ln \alpha_2}\right|\right\}.
\end{equation}
The tuning is then $\Delta_{\text{Spontaneous}}^{-1}$.\footnote{As in \cite{Craig:2013fga}, we do not consider variations with respect to $\mu_1^2$, $\alpha_1$ and $\lambda_1$ and consider them to be fixed. Variations with respect to these parameters lead to slightly larger tuning, which is a consequence of $B_\mu^\text{max}$ having cubic dependence on $f_1$. Our measure of tuning instead measures how close $B_\mu$ must be taken to $B_\mu^\text{max}$ and how much the parameters of $H_2$ are adjusted with respect to those of $H_1$.} The number of parameters in the model is 7 ($\mu_1^2$, $\mu_2^2$, $\lambda_1$, $\lambda_2$, $\alpha_1$, $\alpha_2$ and $B_\mu$). Three of them can be used to obtain the correct value of $v$ and the Higgs mass, as well as specifying $f_1/v$ or requesting a given tuning. A convenient choice is to use $\mu_1^2$, $\alpha_1$ and $B_\mu$ for this. The free parameters are then $\lambda_1$ and the parameters related to $H_2$ only. For convenience, we show all of the following plots for $\lambda_1=\lambda_2=1$. There are then two parameters left: $\mu_2^2$ and $\alpha_2$. As it makes the relation with the results of section \ref{sSec:USmodel} clearer, we present all contour plots in terms of $\mu_2^2/\mu_1^2$ and $-\alpha_2/\alpha_1$. 

\begin{figure}
        \centering
        \begin{subfigure}[b]{0.5\textwidth}
                \centering
                \includegraphics[width=0.95\linewidth, bb = 0 0 411 424]{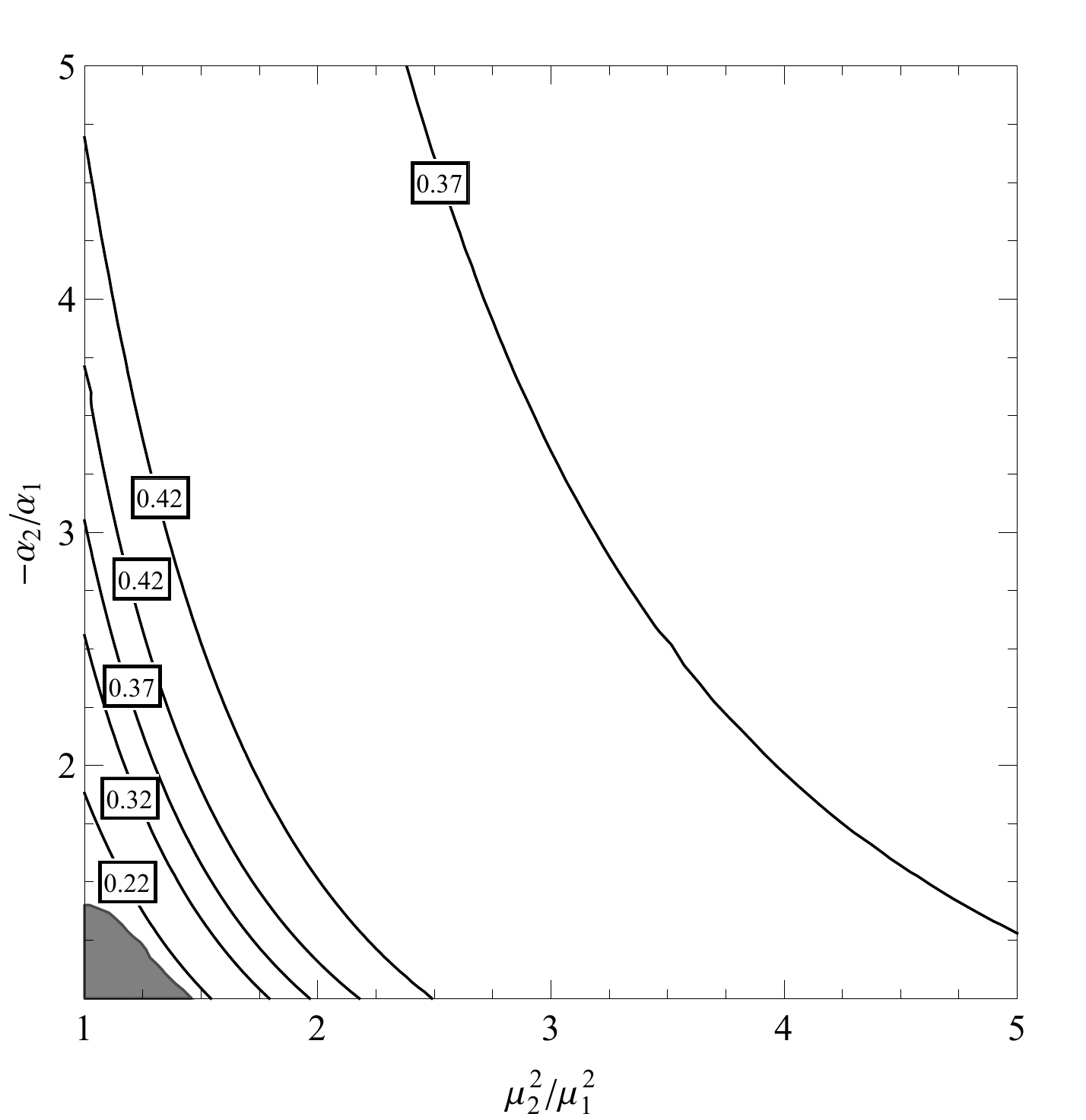}
                \label{Fig:Tuning:A}
        \end{subfigure}%
        ~\quad
        \begin{subfigure}[b]{0.5\textwidth}
                \centering
                \includegraphics[width=0.95\linewidth, bb = 0 0 411 424]{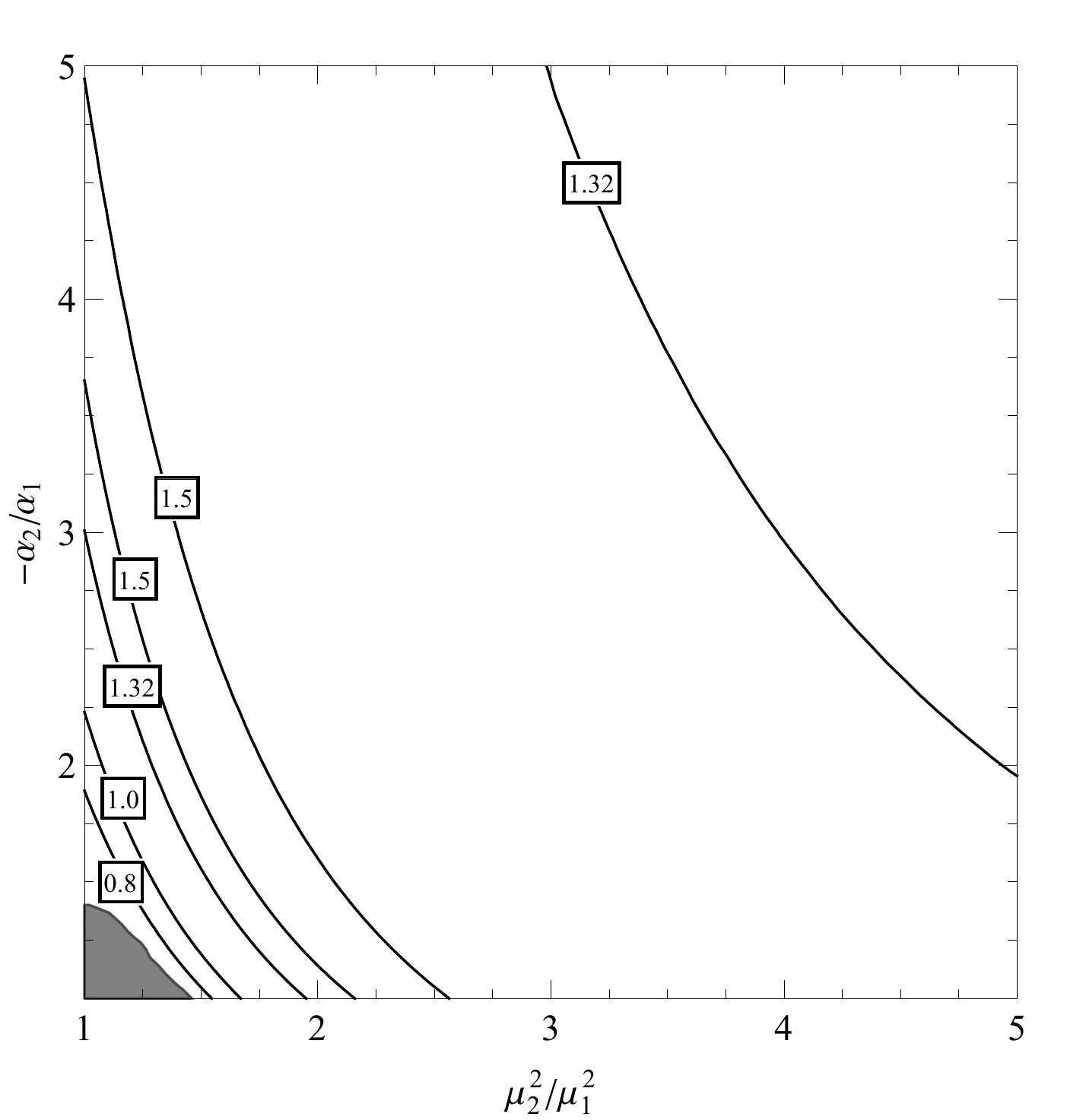}
                \label{Fig:Tuning:B}
        \end{subfigure}
				\caption{Tuning of the spontaneous $\mathbb{Z}_2$ breaking Twin Higgs for a fixed $f_1/v$ of 3. The left panel shows the tuning in percentage and the right one the ratio of the tuning on the Twin Higgs benchmark of 27.7$\%$. The gray area corresponds to the region where the constraints do not accept any solution.}\label{Fig:Tuning}
\end{figure}

\begin{figure}
        \centering
        \begin{subfigure}[b]{0.5\textwidth}
                \centering
                \includegraphics[width=0.95\linewidth, bb = 0 0 383 396]{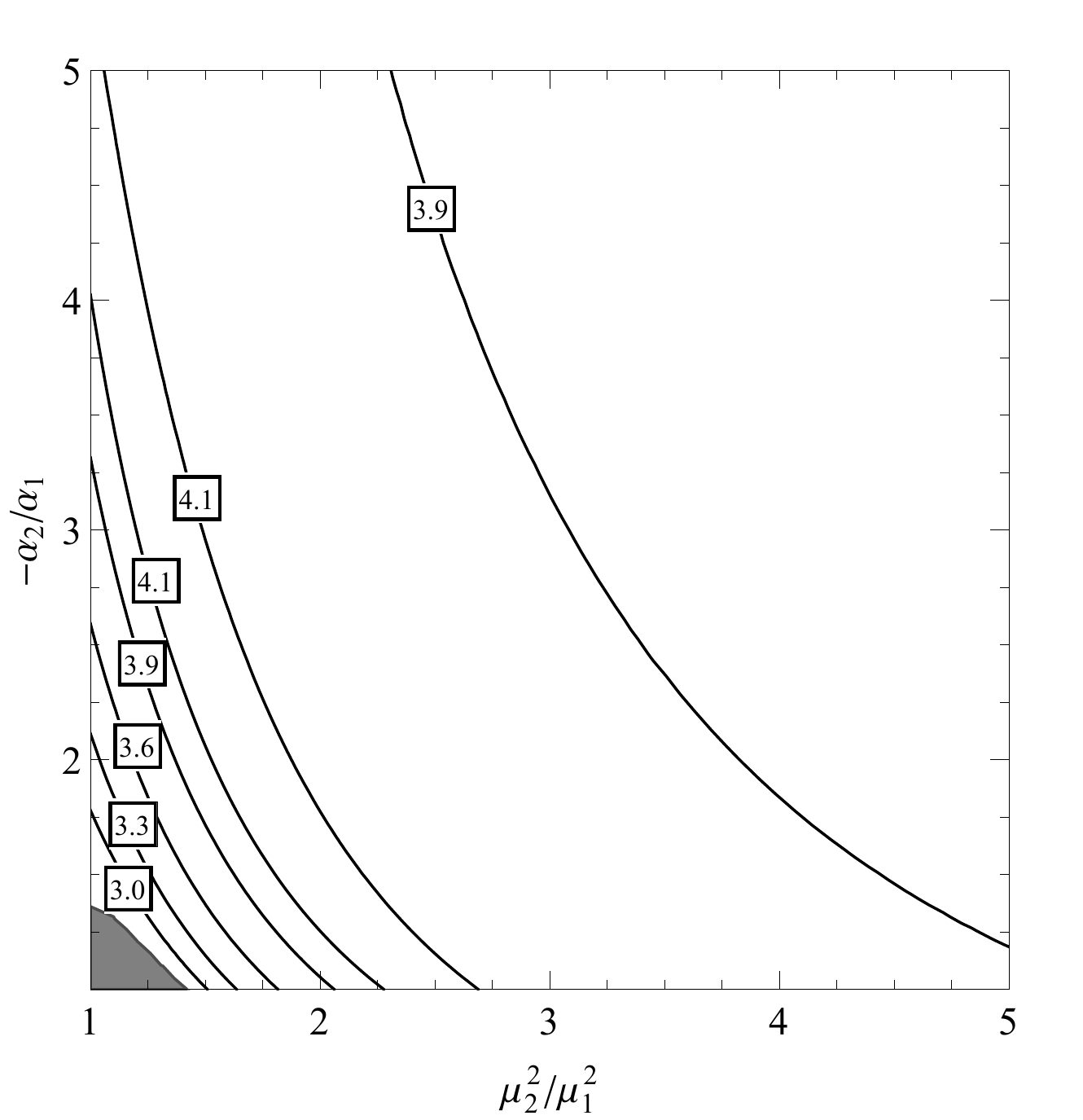}
                \label{Fig:Cutoff:A}
        \end{subfigure}%
        ~\quad
        \begin{subfigure}[b]{0.5\textwidth}
                \centering
                \includegraphics[width=0.95\linewidth, bb = 0 0 383 396]{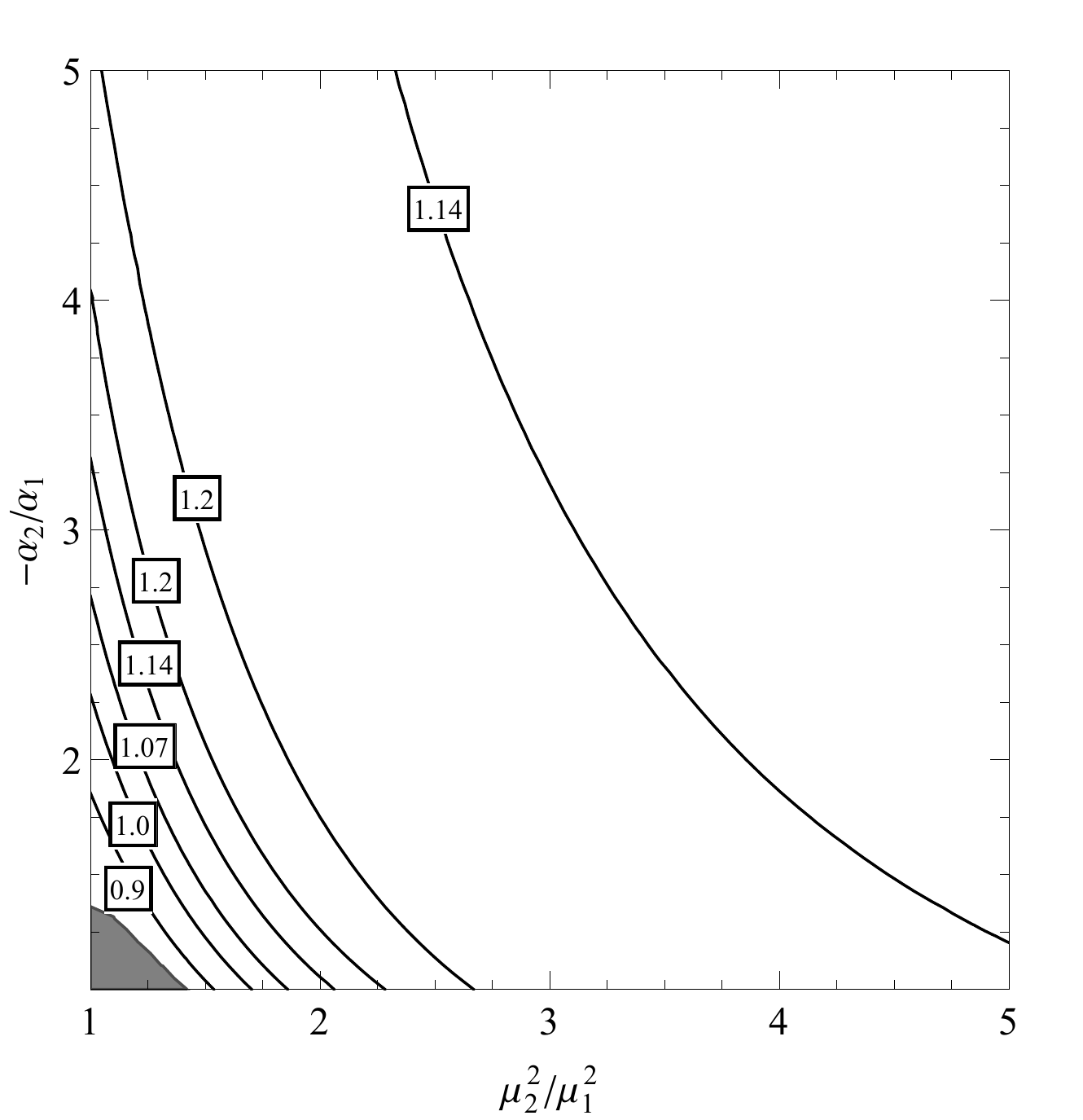}
                \label{Fig:Cutoff:B}
        \end{subfigure}
				\caption{$f_1/v$ for the spontaneous $\mathbb{Z}_2$ breaking Twin Higgs for a fixed tuning of 20$\%$. The left panel shows $f_1/v$ and the right one the ratio of $f_1/v$ on the Twin Higgs benchmark of 3.42. The gray area corresponds to the region where the constraints do not accept any solution.}\label{Fig:Cutoff}
\end{figure}

The left  panel of figure \ref{Fig:Tuning} shows the tuning given a ratio $f_1/v$ of 3. By inspecting \ref{Eq:US:Omega}, one sees that the contour lines correspond roughly to lines of constant $\Omega$. The tuning also approaches a constant as $\Omega$ goes to 0. This corresponds to the behavior expected from the discussion of section \ref{sSec:USmodel}. The gray area corresponds to the region of parameter space where the constraints do not accept any solution. It originates from the impossibility of creating a large enough hierarchy of vevs for $\Omega$ very close to 1. The model is least fine-tuned when $\Omega$ is large enough for feedback to play an important role, while at the same time far away enough from 1 not to be considered fine-tuned. The ratio of the tuning and the corresponding Twin Higgs benchmark of 27.7$\%$ is shown in the right panel of figure \ref{Fig:Tuning}. There is an optimal improvement of 58.1$\%$ and an improvement of 29.2$\%$ in the limit of $\Omega$ going to 0. Conversely, figure \ref{Fig:Cutoff} shows $f_1/v$ for a fixed tuning of 20$\%$. The ratio of $f_1/v$ on the corresponding Twin Higgs benchmark of 3.42 can be seen in the right panel of figure \ref{Fig:Cutoff}. There is an optimal improvement of 22.5$\%$ and an improvement of 12.3$\%$ in the limit of $\Omega$ going to 0. 

Also of interest is the scale at which new physics is expected to become relevant, i.e. the cutoff. New physics is expected where the radiative corrections to the different parameters become large compared to their actual values. The bare parameters must then be tuned against their radiative corrections. The relevant parameters in both Twin Higgs and our model are those that receive quadratic corrections, i.e. the different $\mu^2$'s. An estimate of the cutoff for a fixed tuning can be obtained by taking the value of $\Lambda$ for which the ratio of one of the $\mu^2$'s and its radiative correction drops below said tuning. These corrections are only expected to give an order of magnitude and are roughly given by $|\delta \mu^2| \sim 10\lambda^2 \Lambda^2/(16 \pi^2)$. Using the relations of section \ref{Sec:Model}, the results of $f/v$ can be used to estimate the cutoff. Requesting a tuning of 20$\%$ gives a cutoff of 7.5 TeV for the Twin Higgs. In our model, the cutoff follows a similar pattern to figure \ref{Fig:Cutoff} with an optimal value of 9.2 TeV and a value of 8.4 TeV in the limit of $\Omega$ going to 0.

\section{Concluding remarks}\label{Sec:Conclusion}
In Twin Higgs models, the Higgs is a pseudo-Goldstone of a spontaneously broken approximate $SU(4)$ global symmetry. It is kept light thanks to a $\mathbb{Z}_2$ symmetry that relates the Standard Model sector to a mirror sector. In order for the model to provide a hierarchy between the electroweak scale and the scale of new physics, an explicit $\mathbb{Z}_2$ breaking term is introduced and tuned against the small $SU(4)$ breaking terms.  In this article, we propose a  Twin Higgs model where the $\mathbb{Z}_2$ symmetry is spontaneously broken. It consists of two Higgses that are fundamentals of a global $SU(4)$. When they are decoupled, the vacuum of one of them preserves a $\mathbb{Z}_2$, while the other breaks it spontaneously.  A $B_\mu$-like term that is bilinear in the two Higgses is then introduced. It acts as an effective tadpole and communicates the $\mathbb{Z}_2$ breaking from one sector to the other, resulting in a hierarchy of vevs. This effective tadpole and the feedback between the two Higgses lead to a milder tuning than in the original Twin Higgs.

The phenomenology of the model is quite similar to that of the Twin Higgs. It contains a mirror sector that is not charged under the Standard Model. The two sectors communicate weakly through the Higgs but, as mentioned above, the mixing of the Standard Model Higgs with the $B$-sector is smaller in our model than in the Twin Higgs. On the other hand, this model is a two Higgs doublet model which could lead to additional signatures.

The next logical question concerns a possible UV completion. The obvious guess would be a supersymmetric version of the model. However, SUSY generally leads to a more complicated quartic structure than \ref{Eq:US:H1} and \ref{Eq:US:H2}. This prevents the model from being translated directly to SUSY. In addition, getting the correct sign of the $\alpha_i$'s generally proves to be problematic. The combination of the D-terms and the largest loop corrections provides a negative contribution to the $\alpha_i$ of both the up and down Higgses \cite{Craig:2013fga}. The terms leading to spontaneous $\mathbb{Z}_2$ breaking must therefore originate from the superpotential. One possibility would be to introduce a superpotential term of the form
\begin{equation}\label{Eq:UV:alpha}
	\lambda H_{dA}UH_{dB},
\end{equation}
where $U$ is a fundamental of both $SU(2)_A$ and $SU(2)_B$ and has the appropriate weak hypercharges. Assuming a very large soft mass for $U$ and integrating it out would lead to a positive contribution to $\alpha_d$ and can lead to the correct $\mathbb{Z}_2$ breaking structure.

One other possibility would be to have both $H_u$ and $H_d$ preserve $\mathbb{Z}_2$, but include a NMSSM-like scalar sector that spontaneously breaks $\mathbb{Z}_2$. For example, consider the superpotential
\begin{equation}\label{Eq:US:NMSSM}
	W= \lambda' S' (S_A^2+S_B^2)+\lambda'' S'' S_A S_B
\end{equation}
and assume that both $S'$ and $S''$ have large soft masses and that $S=(S_A,S_B)$ has a negative soft mass squared. The first term preserves a global $O(2)$ symmetry that the second term breaks. Both terms preserve the $\mathbb{Z}_2$ symmetry. However, this symmetry will be broken spontaneously. If $S_A$ couples to the $A$-type Higges and $S_B$ to the $B$-type Higgses, the symmetry breaking is transmitted to the Higgs sector as well. Of course, the viability of these models would require studies of their own.

\acknowledgments
This work was supported in part by the Natural Sciences and Engineering Research Council of Canada (NSERC). HB acknowledges support from the Ontario Graduate Scholarship (OGS). KE acknowledges support from the Alexander Graham Bell Canada Graduate Scholarships-Doctoral Program (CGS D).

\bibliographystyle{JHEP}
\bibliography{Paper1}

\end{document}